\documentclass{aastex63}

\shorttitle{A response to Rubin \& Heitlauf 2019}
\shortauthors{Colin et al.}

\begin{document}

\newcommand{\iap}{\affiliation{CNRS, UPMC, Institut d'Astrophysique de Paris, 98 bis Bld Arago, Paris, France}}
\newcommand{\nbi}{\affiliation{Niels Bohr Institute, University of Copenhagen, Blegdamsvej 17, 2100 Copenhagen, Denmark}}
\newcommand{\uoxford}{\affiliation{Rudolf Peierls Centre for Theoretical Physics, University of Oxford, Parks Road, Oxford OX1 3PU, United Kingdom}}

\author{J. Colin}
\iap

\author[0000-0002-5944-3995]{R. Mohayaee}
\iap

\author[0000-0001-5023-5631]{M. Rameez}
\nbi

\author[0000-0002-3542-858X]{S. Sarkar}
\uoxford

\shortauthors{Colin, Mohayaee, Rameez, Sarkar} 
\title{A response to Rubin \& Heitlauf: ``Is the expansion of the universe accelerating? All signs still point to yes''}

\correspondingauthor{Mohamed Rameez}
\email{Mohamed.Rameez@nbi.ku.dk}

\begin{abstract}
We have shown \citep{Colin:2018ghy} that the acceleration of the Hubble expansion rate inferred from Type Ia supernovae (SNe~Ia) is, at  $3.9\sigma$ significance, a dipole approximately aligned with the CMB dipole, while its monopole component, which can be interpreted as due to a Cosmological Constant or dark energy, is consistent with zero at $1.4\sigma$. This has been challenged by \cite{Rubin:2019ywt} who assert that the dipole arises because we made an incorrect assumption about the SNe~Ia light-curve parameters (viz. took them to be sample- and redshift independent), and did not allow for the motion of the Solar system (w.r.t. the `CMB frame' in which the CMB dipole supposedly vanishes). In fact what has an even larger impact on our finding is that we reversed the inconsistent ``corrections'' made for the peculiar velocities of the SNe~Ia host galaxies w.r.t the CMB frame, which in fact serve to bias the lever arm of the Hubble diagram towards higher inferred values of the monopole. We demonstrate that even if all such corrections are made consistently and both sample- and redshift-dependence is allowed for in the standardisation of supernova light curves, the evidence for isotropic acceleration rises to just $2.8\,\sigma$. Thus the criticism of Rubin \& Heitlauf serves only to highlight the ``corrections'' that must be made to the SNe~Ia data \emph{assuming} the standard $\Lambda$CDM model, in order to recover it from the data. 
\end{abstract}


\section{Introduction} 
\label{sec:intro}

\cite{Rubin:2019ywt} (RH19) make four main criticisms of our paper \citep{Colin:2018ghy} (CMRS19):

\begin{enumerate}

\item RH19 say that we made the ``plainly incorrect assumption'' that the light-curve shape ($x_1$) and colour ($c$) parameters in the  `Spectral Adaptive Lightcurve Template' (SALT2) for SNe~Ia are independent of redshift because \cite{Rubin:2016iqe} (RH16) had demonstrated the need for this. In fact in analysing the `Joint Lightcurve Analysis' (JLA) SNe~Ia catalogue, CMRS19 followed its authors \citet{Betoule:2014frx} who stated that the light curve parameters are redshift-independent. Whereas RH16 advocated that both $x_1$ and $c$ have sample- as well as redshift-dependence, the corresponding increase in parameters (from 10 to 22) is not justified by the Bayesian Information Criterion. 
Moreover \cite{Karpenka:2015vva} has shown  that allowing redshift dependence for $x_1$ and $c$ introduces a large bias in the cosmological parameters inferred using the `Bayesian Hierarchical Model' used by RH16. \cite{Dam:2017xqs} have also emphasised that systematic uncertainties and selection biases in the data need to be corrected for in a model-independent manner \emph{before} fitting to a particular cosmological model.
    
To justify their approach RH19 use a na\"ive Bayesian Information Criterion that penalises each additional parameter with a factor of $\ln(740) \sim 6.6$, where 740 is the number of SNe~Ia in the JLA catalogue. However, since SNe~Ia data are multidimensional (each of the supernovae that make up the sample of 740 are characterised by magnitude $m$ and redshift $z$, in additions to $x_1$ and $c$), any parameters added to the treatment of dimensionality of the datasets must be penalised by \emph{additional} factors scaled up by a minimum of $6 \times \ln(740) \sim 39.6$~\citep{GaoCarroll}. Hence we stand by our statement that the 12 additional parameters added by RH16 and RH19 are not justified by any meaningful Bayesian Information Criterion. (Such a penalty however does not apply to the cosmological parameters $q_\textrm{d}$ and its scale $S$, which are motivated independently by physical considerations.)

 \item The second issue RH19 highlight is that CMRS19 use heliocentric redshifts rather than correcting for the motion of the Solar system, as is \emph{inferred} from the kinematic interpretation of the CMB dipole, to boost to the `CMB frame' in which the universe would supposedly look isotropic. In fact both the discovery papers of \citet{Perlmutter:1998np} and \citet{Riess:1998cb}, as well as all supernova cosmology analyses until \cite{Conley:2011ku}, used \emph{heliocentric} redshifts and made \emph{no} corrections for peculiar velocities. The evidence for cosmic acceleration was based on the observation that: ``\emph{\ldots compared to low redshift SNe~Ia, high redshift ones are on average fainter by $\sim0.15$~mag ($\sim15\%$ in flux) than would be expected in a $\Lambda = 0$ universe}" \citep{Perlmutter:1998np}, however both \citet{Perlmutter:1998np} and \citet{Riess:1998cb} employed SNe~Ia down to $z=0.01$ where, as RH19 correctly point out, peculiar velocities have an impact as large as $\sim 0.27$ mag.
 
 We are in fact motivated to look for a dipole in the deceleration parameter motivated by the covariant argument of \cite{Tsagas:2011wq} and \cite{Tsagas:2015mua} that `tilted' observers such as ourselves who are embedded in a bulk flow can be misled into inferring cosmic acceleration along the direction of that bulk flow. Accordingly we must look at the data as measured and not make ``corrections'' as RH19 do that necessarily assume an underlying FLRW model (wherein such deep bulk flows would in fact \emph{not} exist).

 RH19 assert further that when the full JLA peculiar-velocity covariance matrix is included in the analysis, they find no statistically significant anisotropy at 2$\sigma$, even when using heliocentric redshifts. This statement is trivially true and simply reflects how uncertainties behave. Adding additional uncertainties will weaken the statistical significance of hypothesis tests, in particular the peculiar velocity component of the covariance matrix adds additional uncertainties to the SNe~Ia at $z<0.06$. The covariant effects of peculiar velocities are, for us, a signal --- not a source of uncertainty.

RH19 also comment that our justification of the heliocentric frame choice is ``weak''. Apart from the fact that the primary hypothesis we set out to test requires us to consider the actually measured redshifts, we wish to point out that both \citet{Hudson:2004et} and \citet{Carrick:2015xza} (the flow models usually used for peculiar velocity corrections) provide evidence for bulk flows of 372 $\pm$ 127 km s$^{-1}$ and 159 $\pm$ 23 km s$^{-1}$ respectively, originating from structure beyond the scale of these models. Thus bulk flows \emph{definitely} exist on larger scales, according to the very models used to carry out the ``corrections'' that RH19 espouse.

Moreover the usual interpretation of the CMB dipole as being of kinematic origin is questioned by the observation of significantly larger dipoles in other surveys of sources at much \emph{smaller} redshift~\citep[see e.g.][]{Singal:2011dy,Rubart:2013tx,Kothari:2013gya,Colin:2017juj,Rameez:2017euv,Bengaly:2017slg}. Observing the expected parallax and modulation of higher multipoles in the CMB anisotropy would settle the issue, however even the Planck data do not yield a convincing detection of such effects \citep[see Table 1 of ][]{Agahanim:2013suk}.

\item The third criticism by RH19 is that we ignore cosmological results from surveys carried out in the Southern hemisphere such as the Carnegie Supernova Project \citep{Krisciunas:2017yoe} and the Dark Energy Survey \citep{Abbott:2018wog}. \cite{Krisciunas:2017yoe} present no cosmological results, while \citep{Abbott:2018wog} employ peculiar velocity ``corrections'' based on \citet{Carrick:2015xza} just as \cite{Scolnic:2017caz} did for Pantheon. Neither of these datasets are publicly available in a form that we can use, unlike the JLA catalogue which provides full details of the lightcurve fitting parameters and all individual covariances. 

RH19 also question why we did not find a correlation between the dipole and the monopole. In fact we \emph{do} --- as seen in Figure 3 of CMRS19, the monopole $q_\textrm{m}$ prefers more negative values as the dipole $q_\textrm{d}$ approaches zero.

\item Finally RH19 say that our model for the dipole anisotropy is pathological when modeling an isotropic universe since no lower limit is set on the scale $S$ of the exponential fall-off of $q_\textrm{d}$. In fact our code did include a lower bound on $S$ of 0.01 (this ought to have been stated in CMRS19). The ``pathology'' alleged by RH19 never arises, other than within the Monte Carlo datasets of the isotropic toy Universe which they investigate in this context.

\end{enumerate}

We therefore thank RH19 for reasserting our points and for vividly illustrating in their Figures~2 and 3 that any evidence for cosmic acceleration from SNe~Ia is explicitly dependent on how peculiar velocities (and related additional uncertainties) are treated. As a guide to the coloured contours in their figure the reader may wish to note that:

\begin{itemize}

\item Pink: is the same as CMRS19 and uses heliocentric redshifts, as in all supernova cosmology analyses starting with \cite{Riess:1998cb} and \cite{Perlmutter:1998np}, up until \cite{Conley:2011ku}. This confirms our results.

\item Green: uses heliocentric redshifts and assumes sample and redshift dependence of $x_1$ and $c$. In their Figure~3., the peculiar velocity ``correction'' covariance matrix is added, treating the peculiar flow as distinct from the Hubble flow, i.e. uncertainties assigned to the $z<0.06$ SNe~Ia  observables are artificially enlarged.

\item Blue: Same as Green, but redshifts are now corrected to the `CMB frame', assuming that the CMB dipole is of kinematic origin (even though this is questionable given the \emph{larger} dipole seen in the rest frame of radio galaxies).

\item Grey: In addition to the assumptions in Blue, the $z<0.06$ redshifts are ``corrected'' for the motion of the SNe~Ia with respect to the CMB rest frame. Since the velocities beyond the scale of the flow model are unknown but these models admit a residual bulk flow, the redshifts now encode a model of the Universe in which a local spherical volume of $\sim$200 Mpc radius is \emph{smashing} into the rest of the Universe, which is treated as at rest.

\end{itemize}

\subsection{The accelerated expansion of the Universe}

The earliest claims of an accelerating Universe ~\citep{Perlmutter:1998np, Riess:1998cb} employed 42 and 50 SNe~Ia respectively. Subsequent tests on larger datasets have concluded that the evidence is strongly dependent on the assumption of isotropy \citep{Seikel:2008ms}, while supernova datasets themselves show evidence for a bulk flow i.e. \emph{anisotropy} \citep{Colin:2010ds,Feindt:2013pma}.  

\citet{Nielsen:2015pga} found using a  Maximum likelihood estimator that even with the 740 SNe~Ia in the JLA catalogue, the evidence for acceleration is marginal ($<3\sigma$). This analysis employed the JLA catalogue as it ships, i.e. with peculiar velocity corrections. Since the peculiar velocity corrections introduce an arbitrary discontinuity within the data and modify the low redshift SNe~Ia which serve to fix the lever arm of the Hubble diagram, the impact of these corrections on the evidence for cosmic acceleration is worth examining. Note that corrections for peculiar velocities (PV) in SNe~Ia analyses were first introduced in \cite{Conley:2011ku} who also first used CMB frame redshifts.

The peak $B$-band apparent magnitudes $m_B^*$ of the SNe~Ia as reported in the JLA catalogue \citep{Betoule:2014frx} is related to the distance modulus as:
\begin{equation}
\label{eq:DMOD}   
\mu_\mathrm{SN} = m_B^* - M^0_B + \alpha x_1 - \beta c ,
\end{equation}
in terms of the absolute magnitude $M^0_B$ and the stretch and colour corrections, respectively $x_1$ and $c$, applied to the SNe~Ia lightcurves using the SALT2 template. In the standard Friedman-Lem\^aitre-Robertson-Walker (FLRW) framework, this is related to the luminosity distance $d_\mathrm{L}$ as:
\begin{eqnarray}
& \mu &\equiv 25 + 5 \log_{10}(d_\mathrm{L}/\mathrm{Mpc}), 
 \quad \mathrm{where:} \nonumber \\
& d_\mathrm{L} &= (1 + z_\mathrm{hel}) \frac{d_\mathrm{H}}{\sqrt{\Omega_k}} 
 \mathrm{sin}\left(\sqrt{\Omega_k} \int_0^{z_{cmb}} \frac{H_0 \mathrm{d}z'}{H(z')}\right), \mathrm{for \  } \Omega_k > 0
 \nonumber \\
&  &= (1 + z_\mathrm{hel}) d_\mathrm{H} \int_0^{z_{cmb}} \frac{H_0 \mathrm{d}z'}{H(z')}, \mathrm{for \  } \Omega_k = 0
 \nonumber \\
&  &= (1 + z_\mathrm{hel}) \frac{d_\mathrm{H}}{\sqrt{\Omega_k}} 
 \mathrm{sinh}\left(\sqrt{\Omega_k} \int_0^{z_{cmb}} \frac{H_0 \mathrm{d}z'}{H(z')}\right), \mathrm{for \  } \Omega_k < 0
 \nonumber \\
& d_\mathrm{H} &= c/H_0, \quad H_0 \equiv 
 100h~\mathrm{km}\,\mathrm{s}^{-1}\mathrm{Mpc}^{-1}, 
 H = H_0 \sqrt{\Omega_\mathrm{m} (1 + z)^3 + \Omega_k (1 + z)^2  + \Omega_\Lambda}.
\label{DLEQ}
\end{eqnarray}
Here $d_\mathrm{H}$ is the Hubble distance and $H$ the Hubble parameter ($H_0$ being its present value), and $\Omega_\mathrm{m} \equiv \rho_\mathrm{m}/(3H_0^2/8\pi G_\mathrm{N}), \Omega_\Lambda \equiv \Lambda/3H_0^2, \Omega_k \equiv -kc^2/H^2_0 a^2_0$ are the matter, cosmological constant and curvature densities in units of the critical density. In the FLRW framework these are related by the `cosmic sum rule': $1=\Omega_\mathrm{m} + \Omega_\Lambda + \Omega_k$.  

For the redshifts $z \lesssim 1.4$ to be considered, the luminosity distance $d_\mathrm{L}$ can also be quite accurately, to within 7\% of eq.(\ref{DLEQ}), expanded as a Taylor series in terms of the Hubble parameter $H_0$, the deceleration parameter $q_0 \equiv -\ddot{a} a/\dot{a}^2$, and the jerk $j_0 \equiv \dot{\ddot{a}}/aH^3$. This is `cosmography', i.e. independent of assumptions about the content of the universe, and in particular not subject to the above `sum rule' (which holds for the standard $\Lambda$CDM model in which $q_0 \equiv \Omega_\mathrm{m}/2 - \Omega_\Lambda$). Further modified to explicitly show the dependence on the measured heliocentric redshifts, this writes (RH19):
\begin{equation}
\label{eq:Qkin}    
d_\mathrm{L} (z, z_\mathrm{hel}) = \frac{cz}{H_0} \left[1 + \frac{1}{2}(1 - q_0)z - \frac{1}{6} (1-q_0 - 3q^2_0 + j_0 - \Omega_k) z^2 \right] \times \frac{1+z_\mathrm{hel}}{1+z} ,
\end{equation}
where $z$ can be the measured heliocentric redshift, boosted to the CMB frame, or boosted to the CMB frame with further peculiar velocity ``corrections'' applied.

\begin{figure}
\begin{center}
\includegraphics[scale=0.35,angle=0]{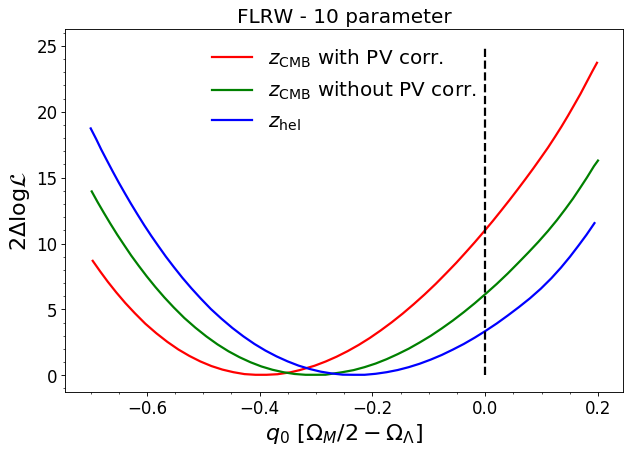}\includegraphics[scale=0.35,angle=0]{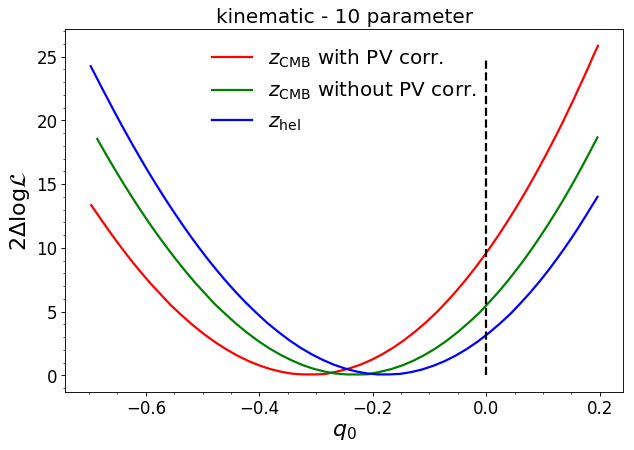}
\caption{Left: The profile likelihood for the FLRW analysis, following \citet{Betoule:2014frx} for the colour $c$ and stretch $x_1$ corrections (corresponding to Table~\ref{tab:llhfitparsflrw}). Right: The same for the kinematic analysis (corresponding to Table~\ref{tab:llhfitparskinem}).}
\label{fig:proflikescan}
\end{center}
\end{figure}

\begin{figure}
\begin{center}\includegraphics[scale=0.35,angle=0]{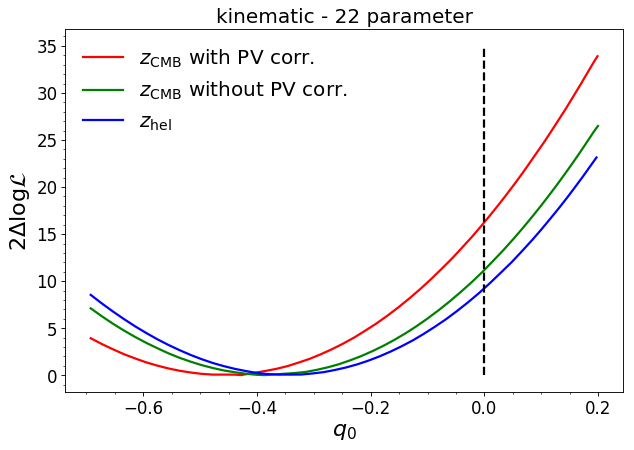}
\caption{The profile likelihood for the 22 parameter kinematic analysis of \cite{Rubin:2019ywt}, employing the sample- and redshift-dependent treatment of $c_0$ and $x_{1,0}$ advocated by \cite{Rubin:2016iqe} (corresponding to Table~\ref{tab:llhfitparskinemRH2}).}
\label{fig:proflikescan22pars}
\end{center}
\end{figure}

We redo the analysis of \citet{Nielsen:2015pga} in three different ways and present the results in Figures \ref{fig:proflikescan} and \ref{fig:proflikescan22pars}, and in Tables \ref{tab:llhfitparskinem} and \ref{tab:llhfitparsflrw}, respectively for the kinematic Taylor expansion (eq. \ref{eq:Qkin}) and the standard $\Lambda$CDM model (eq. \ref{DLEQ}). For each case we also show the fit quality when $q_0$ is held at zero (``No accn.").

\begin{enumerate}

\item $z_\mathrm{CMB}$ with PV corr.: This employs the data exactly as in \citet{Nielsen:2015pga}. The CMB frame redshifts are used, with further corrections made for the peculiar velocities of the SNe~Ia w.r.t. the CMB frame, and the peculiar velocity covariance matrix is included.

\item $z_\mathrm{CMB}$ without PV corr.:  Now CMB frame redshifts are used \emph{without} correcting for the flow of the SNe~Ia w.r.t. this frame and the peculiar velocity component of the covariance matrix is excluded. Note that transforming from heliocentric to CMB frame redshifts still requires assuming that the CMB dipole is kinematic in origin. 

\item $z_\mathrm{hel}$: Finally heliocentric redshifts are used, no corrections are employed and the peculiar velocity component of the covariance matrix is excluded. This is just what was done by \cite{Perlmutter:1998np} and \cite{Riess:1998cb}, as well as in all supernova cosmology papers until \cite{Conley:2011ku}.

\end{enumerate}

Our results in Tables \ref{tab:llhfitparsflrw} and \ref{tab:llhfitparskinem} illustrate that the peculiar velocity corrections serve to bias the data towards \emph{higher} acceleration (more negative $q_0$). Using heliocentric observables, as employed by \cite{Perlmutter:1998np} and \cite{Riess:1998cb} as well as all supernova cosmology analyses before \cite{Conley:2011ku}, the change in $2\mathrm{log}\mathcal{L}$ between the best-fit model and the one with zero acceleration is only 3.3, indicating that the preference for acceleration is $<1.4\sigma$. 

\begin{table*}
\caption{Results of a 10-parameter fit using eq.(\ref{DLEQ}), following \citet{Betoule:2014frx} for the colour $c$ and stretch $x_1$ corrections.}
\begin{center}
\label{tab:llhfitparsflrw}
\begin{tabular}
{| l | c | c | c | c | c | c | c | c | c | c | c | c|}
\hline
 Fit & -2log$\mathcal{L}_\mathrm{max}$ & $\Omega_\mathrm{m}$ & $\Omega_{\Lambda}$ & $\alpha$ & $x_{1,0}$ & $\sigma_{x_{1,0}}$ & $\beta$ & $c_0$  & $\sigma_{c_0}$ & $M_0$ & $\sigma_{M_0}$ \\ 
\hline\hline
1. $z_\mathrm{CMB}$ w. PV corr.  &-221.9&0.3402 &0.5653&0.1334&0.03849&0.9321&3.056&-0.01584&0.07107&-19.05&0.1073 \\
As above + No accn. & -211.0 &0.0699&0.03495&0.1313&0.03275&0.9322&3.042&-0.01318&0.07104&-19.01&0.1087 \\
\hline\hline
2. $z_\mathrm{CMB}$ w/o PV corr. &-210.5&0.2828&0.4452&0.135&0.0389&0.9315&3.024&-0.01686&0.07109&-19.04&0.11 \\
As above + No accn. &-204.4&0.07345&0.03673&0.1334&0.03451&0.9316&3.013&-0.01491&0.07105&-19&0.1109\\
\hline\hline
3. $z_\mathrm{hel}$, no PV&-198.2&0.2184&0.3387&0.1333&0.03974&0.9317&3.018&-0.01448&0.07111&-19.03&0.1116\\
As above + No accn. &-194.9&0.05899&0.0295&0.1322&0.03649&0.9317&3.006&-0.01302&0.07108&-19&0.1122 \\
\hline
\end{tabular}
\end{center}
\end{table*}

\begin{table}
\caption{Results of a 10-parameter fit using eq.(\ref{eq:Qkin}), following \citet{Betoule:2014frx} for the colour $c$ and stretch $x_1$ corrections.}
\begin{center}
\label{tab:llhfitparskinem}
\begin{tabular}
{| l | c | c | c | c | c | c | c | c | c | c | c | c|}
\hline
 Fit & -2log$\mathcal{L}_\mathrm{max}$ & $q_0$ & $j_0 - \Omega_k$ & $\alpha$ & $x_{1,0}$ & $\sigma_{x_{1,0}}$ & $\beta$ & $c_0$  & $\sigma_{c_0}$ & $M_0$ & $\sigma_{M_0}$ \\ 
\hline\hline
1. $z_\mathrm{CMB}$ w. PV corr. &-220.8&-0.311&0.02946&0.1332&0.03809&0.9323&3.056&-0.01591&0.07108&-19.05&0.1074 \\
 As above + No accn. & -211.4&0&-0.8211&0.1312&0.03267&0.9321&3.041&-0.01295&0.07102&-19.01&0.1087 \\
\hline\hline
2. $z_\mathrm{CMB}$ w/o PV corr. &-210.1&-0.2332&-0.2328&0.1348&0.03859&0.9317&3.023&-0.01689&0.07109&-19.03&0.11\\
 As above + No accn. &-204.8&0&-0.8183&0.1333&0.03446&0.9315&3.012&-0.01473&0.07104&-19&0.1108\\
\hline\hline
3. $z_\mathrm{hel}$, no PV &-198.1 &-0.1764&-0.4405&0.1332&0.03955&0.9318&3.017&-0.01449&0.07111&-19.03&0.1116 \\
As above + No accn. &-195.1 &0&-0.8534&0.1321&0.03645&0.9317&3.006&-0.01288&0.07106&-19&0.1122 \\
\hline
\end{tabular}
\end{center}
\end{table}

Table~\ref{tab:llhfitparskinemRH2} shows that even if we adopt the sample- and redshift-dependent treatment of $c_0$ and $x_{1,0}$ advocated by RH16 and RH19, \emph{and} use redshifts transformed to the CMB frame, the evidence for acceleration remains $<3\sigma$ unless further arbitrary ``corrections'' are also made for the flow of the SNe~Ia with respect to the CMB frame . 

\begin{table*}
\caption{Selected parameters from the 22-parameter fit using eq.(\ref{eq:Qkin}) for the luminosity distance and the sample- and redshift-dependent treatment of colour $c$ and stretch $x_{1}$ distributions adopted by \cite{Rubin:2016iqe,Rubin:2019ywt}.}
\label{tab:llhfitparskinemRH2}
\begin{tabular}{| l | c | c | c | c | c | c | c |}
\hline
Fit & -2log$\mathcal{L}_\mathrm{max}$ & $q_0$ & $j_0 - \Omega_k$  & $\alpha$ &  $\beta$ & $M_0$ & $\sigma_{M_0}$ \\ 
\hline\hline
1. $z_\mathrm{CMB}$& -339.5&-0.4577&0.1494&0.1334&3.065&-19.07&0.1065 \\
As above + No accn. &-323.3&0&-1.349&0.1312&3.046&-19.01&0.108 \\
\hline\hline
2. $z_\mathrm{CMB}$ w/o PV corr &-328.3&-0.3776&-0.1781&0.1349&3.034&-19.06&0.1091 \\
As above + No accn. &-317.2&0&-1.326&0.1331&3.017&-19.01&0.1102\\
\hline\hline
3. $z_\mathrm{hel}$ &-316.1&-0.3448&-0.3651&0.1333&3.027&-19.05&0.1105\\
As above +  No accn. &-306.9&0&-1.378&0.1317&3.005&-19.01&0.1117 \\
\hline
\end{tabular}
\end{table*}

We leave it to the reader to judge whether it is appropriate to make ``corrections''  for peculiar velocities but leave the supernovae immediately outside the flow volume uncorrected, noting that such a procedure induces a directional bias (which in the JLA uncertainty budget is simply assigned an uncorrelated variance of $c\sigma_z = 150$~km~$s^{-1}$). An isotropic acceleration can be extracted from the data only by ``correcting'' over half of all observed supernovae to the CMB frame --- to which convergence has never been observationally demonstrated. 

In fact the peculiar velocity ``corrections'' affect  the lever arm of the Hubble diagram in a non-obvious manner. The subsequent Pantheon compilation \citep{Scolnic:2017caz} initially included peculiar velocity corrections far beyond the extent of any known flow model of the Universe.\footnote{\href{https://github.com/dscolnic/Pantheon/issues/2}{https://github.com/dscolnic/Pantheon/issues/2}} When this bug was fixed, both the magnitudes and heliocentric redshifts of the corresponding supernovae were found to be discrepant \citep{Rameez:2019wdt}.\footnote{\href{https://github.com/dscolnic/Pantheon/issues/3}{https://github.com/dscolnic/Pantheon/issues/3}.}

\section{Conclusions}

As demonstrated by CMRS19 and confirmed by RH19, the  acceleration of the Hubble expansion rate inferred from the SNe~Ia magnitudes and redshifts as measured (in the heliocentric frame) is described by a dipole anisotropy. To infer an isotropic acceleration i.e. a monopole (such as can be attributed to a Cosmological Constant or dark energy), the supernovae must have their redshifts boosted to the (possibly mythical) CMB rest frame, and the redshifts of the low-$z$ objects ``corrected'' further for their motion w.r.t. the CMB frame. This is done using bulk flow models \citep{Hudson:2004et,Carrick:2015xza} which already \emph{assume} that the universe is well-described by the standard $\Lambda$CDM model. Moreover to boost the significance of the monopole acceleration above $\sim3\sigma$ the parameters describing the supernova light curves need to be empirically modelled \emph{a posteriori} as being both sample- and redshift-dependent --- violating both basic principles of unbiased hypothesis testing and the Bayesian Information Criterion. 

Whether dark energy is just a manifestation of inhomogeneities is a topic of fierce debate \citep{Buchert:2015iva}. It is necessary to carefully scrutinise which reference frame the supernova data have been ``corrected'' to before drawing important conclusions. Whereas there is only one relevant frame in the FLRW model, in the actual inhomogeneous Universe (which can only be \emph{approximately} represented by a FLRW model), observers have a broad choice of ``corrections'' they can make. In this regard fitting a Hubble diagram of \emph{heliocentric} observables, as was done in all supernova cosmology analyses until 2011, and employing peculiar velocity corrections after transforming to the CMB frame as was done subsequently (from \cite{Conley:2011ku} onwards), simply amount to different choices of corresponding 2-spheres within the `null fitting' procedure described by \cite{Ellis:1987zz}. The cosmological `fitting problem' these authors highlight is thus at the heart of the present debate.

\end{document}